# Mapping the Probability of Microlensing Detection of Extra-Solar Planets

Karan Molaverdikhani[1][†], Maryam Tabeshian[2]
[1]M.Sc. in Aerospace Eng., Sharif University of Technology, Iran
[2]B.A in Physics, City University of New York, Hunter College, USA

**ABSTRACT**

The growing rate of increase in the number of the discovered extra-solar planets which has consequently raised the enthusiasm to explore the universe in hope of finding earth-like planets has resulted in the wide use of Gravitational Microlensing as a planet detection method. However, until November 2009, only 9 out of the overall 405 discovered exoplanets have been detected through Microlensing, a fact which shows that this method is relatively new in the detection of extra-solar planets. Therefore, preparing a map of the sky which pinpoints the regions with higher probability of planet detection by this method and is drawn based on the available equipments and other regional factors would, indeed, help speed up the discovery of exoplanets. This paper provides calculations and reasoning to suggest looking toward two distinct regions in constellations Centaurus and Sagittarius in addition to the customary Galactic Bulge in the search for other habitable worlds.

## 1 INTRODUCTION

Our understanding of planetary systems in the past two decades has gone beyond the limits of our Solar System. With more than 405* extrasolar planets found so far, we have moved one step closer to finding the answer to the ultimate question: Are we alone in the Universe? Moreover, this new image of other planetary systems has proved that ours is nothing special. Although planetary systems have been found that somewhat resemble our own, we have also seen profound differences and a variety out there.

Once part of popular science fictions, discovery of other Earths has now been transformed into reality. However, since most detection techniques are sensitive to bigger planets closer to their parent stars, Hot Jupiters have been found more frequently, with fewer terrestrial planets seen in the list. Out of the five popular planet detection techniques, Gravitational Microlensing is the most capable of detecting rocky, earth-sized planets and seems to be more promising in the search for other habitable worlds.

### 1.1 OBJECTIVES

The objective of this paper is to draw a map of the Milky Way Galaxy upon which one may find regions of the sky with higher probability of detecting earth-like planets. Our scientific and technical capability is of course important in the detection of such planets; however, technological readiness is not the overall concern of this paper as it only aims at providing a map of the Milky Way with highest concentration of rocky planets in the habitable zones of their stars. Therefore, the present work hopes to attain the following objectives:

- o  To provide the astronomical information needed for current and future Microlensing search;
- o  To specify the most promising regions of the Galaxy where the search for earth-like planets should be focused on;
- o  To leave out the volume of space that does not seem to yield as many habitable worlds;
- o  And to characterize the Microlensing target stars.

[†] E-mail: k.molaverdi@gmail.com
* As of 13 November 2009, according to the latest statistics by Paris Observatory

With these objectives in mind, the primary goal of this work is to provide a map showing detection probability of stars throughout the Milky Way Galaxy that fall in the Galactic Habitable Zone and could potentially host habitable planets.

With NASA's Kepler Spacecraft launched in 2009 which does provide new data on exoplanets, results of this study will further influence future Microlensing research and will lay the foundation for the next decade of studies relevant to the search for life on other worlds. On the other hand, getting to know where to look for in search for earth-sized planes will determine the scope of future ground-based and space missions and clear many of their key design parameters. Furthermore, finding other systems like our own is critical in our understanding of how planetary systems evolved.

This document is intended to be inclusive of the search for extrasolar planets using planetary Microlensing, the only method among others that is capable of detecting smaller terrestrial planets.

## 1.2 PLANETARY DETECTION METHODS: A BRIEF OVERVIEW

Ever since the first planet outside the Solar System was found in the early 1990s, a number of planetary detection methods have been proposed, with five being more practiced today. In order of their popularity, these methods include:

- o  Radial Velocity/Wobble Method
- o  Photometric Transit Method
- o  Planetary Microlensing
- o  Pulse Method
- o  Eclipsing Binary Method

The most popular and widely practiced technique to date, the Radial Velocity Method, measures the slight shift or wobble in the position of a star which is due to the gravitational effect of the planet(s) around it. The Transit Method detects the dip in the starlight as a result of a planet crossing in front of it. This transit causes the starlight to change by 0.01 to 1 percent, according to the relative diameter of the star and its planet.

The Planetary Microlensing technique makes use of the Einstein's idea of curvature of space-time by studying the small change in the lensing effect of an intervening star on a background source due to the presence of an accompanying planet.

Variations in the arrival times of pulses from rapidly spinning neutron stars suggest presence of planets around them in a technique called the Pulse Method.

And finally, in case of binary stars, perturbation in the periods of eclipses can be taken as an indicator of the existence of planet(s) around them that is studied in the Eclipsing Binary Method.

Each of the above-mentioned methods has its own advantages and drawbacks. Although planetary Microlensing can find small planets, Microlensing events are rare and essentially non-repeatable, which suggests close-in collaboration between several groups across the globe to detect the planet when the event occurs.

## 2  DEVELOPMENT OF IDEAS & ORGANIZATION OF THOUGHTS

In addition to a critical assessment of the technology needed for obtaining an accurate light curve for each Microlensing event, which is not the subject of this study, the following criteria were examined in order to better estimate the fraction of stars with terrestrial-sized, potentially habitable planets through Gravitational Microlensing:

1. An assessment of the Lensing parameters as apply to the Microlensing detection of exoplanets including:

-Mass ratio of the lensing star and its planet, Orbital eccentricity, Planet's semi-major axis, Stellar spectral type, Inclination of planet's plane of orbit to the observer's line of sight, Density of the Microlensing tube, Ratio of the star to planet radii, Stellar phase, Source flux, Event rate

2.  Stellar metallicity
3.  Galactic Habitable Zone
4.  The volume of space that should be studied for a Microlensing event. This was attained by considering the following items:

-Relative velocities of the observer, lens, and the source

-Distribution of mass in the Milky Way Galaxy to find regions with highest stellar or interstellar mass density

5.  The amount of dust in the habitable zone of the target star and its effect on the gravitational influence of the planet which could ultimately result in miscalculation of the planetary mass
6.  The wavelength range, or spectral bandpass, that is necessary to detect stars with earth-like planets
7.  The frequency of earths in each planetary system and factors that affect formation of rocky planets: i.e. stellar spectral type, distance between Jupiter-like planets to their star, etc. This will include detections of gas giant planets to determine their disaster zones by a variety of techniques (including perhaps transit detections of several earth-mass planets from COROT, MOST, Kepler and future the Space Interferometry Mission or SIM Planet Quest as well as ground-based observatories)

8. A revision of what should be called "Habitable Zone" for each type of star

9. Regions of the sky where Microlensing events are seen more frequently Moreover, as this map is being developed with the development of technology and its consequent effect on the advancement of our knowledge, it is necessary to develop the scientific community to prioritize the target list to identify stars most likely to harbor habitable exoplanets.

## 2.1 DERIVING THE FORMULAE

For a background point source, the amplification due to a point mass lens is given as:

$$A(u) = \frac{u^2 + 2}{u\sqrt{u^2 + 4}} \quad (1)$$

where u(t) is the impact parameter, the projected angular separation between lens and source, measured in units of $\theta_E$, the angular Einstein radius of the lens, defined as:

$$\theta_E = \sqrt{\frac{D_{LS}}{D_S D_L} \frac{4GM}{c^2}} \quad (2)$$

where $D_S$ and $D_L$, are distances from the observer to the source and the foreground lens respectively and $D_{LS}$ is the distance between the lens to the source, assuming the observer, lens, and the source to be all stationary. However, since the three points are moving relative to one another, we may then define x as the ratio of the distances between the observer to the lens and the source ($x = D_L/D_S$), where $0 \leq x \leq 1$, and rewrite Eq. (2):

$$\theta_E = \sqrt{\frac{(1-x)}{D_S x} \frac{4GM}{c^2}} \quad (3)$$

This corresponds to a linear Einstein Radius ($R_E$) shown in Eq. (4):

$$R_E = \sqrt{D_S x(1-x) \frac{4GM}{c^2}} \quad (4)$$

We may now define the Event Rate $\Gamma$, the rate at which lenses enter the Microlensing Tube, as the product of the number of observed source stars Ns, the number area density of the lens star, the effective transverse velocity v, the cross section $2D_L\theta_E$, and x, provided that we have a Microlensing event with $u < 1$, corresponding to a magnification $A > (3/\sqrt{5}) \approx 1.34$. Therefore, we may now find $\Gamma$ as follows (M. Dominik et al.):

$$\Gamma = \frac{4\sqrt{G}}{c} N_s D_s^{3/2} \times$$
$$\int_0^1 \rho(xD_S)\sqrt{x(1-x)}v(x)dx \times \quad (5)$$
$$\int_0^\infty \sqrt{M}\xi(M)dM$$

where $v_L$, $v_S$, and $v_O$ are the velocities of the lens, source, and observer perpendicular to the line of sight and the effective transverse velocity (v) is defined as:

$$v = D_L\mu = |v_L - xv_S - (1-x)v_O| \quad (6)$$

## 2.2 EFFECTIVE PARAMETERS IN THE CALCULATION OF EVENT RATE

As seen in Equation (5), two parameters affect $\Gamma$ in a Microlensing event: Relative Velocities and Density Distribution. It is important to note that due to symmetry conditions, we only needed to model half of the Galaxy, and then applied the same model to the other half.

### 2.2.1 Relative Velocities

Since we know the rotation curve for the disk of the Milky Way Galaxy, we may determine the velocities of each point source relative to the observer. To calculate these velocities, one must first calculate velocities relative to the Galactic center. These velocities are affected by dark matter or the predominant mass in the Galaxy and can be expressed in a simple formula:

$$v(r) = 25.62\ln(r) + 172.3 \quad (7)$$

where r is the distance between the point source to the center of the Milky Way Galaxy in units of kpc, and v is given in km/s.

Therefore, the first parameters to consider are the velocities of the observer, lens and the source relative to the observer's line of sight, which can be found using Figure (1).

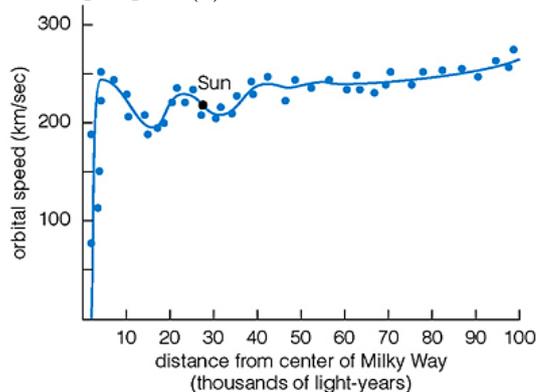

**Figure 1 - The rotation curve for the disk of the Milky Way Galaxy (Scott Schneider 2007)**

The curve in the above graph can be roughly estimated using Equation (7). The uncertainty resulting from this estimation is always less than 20% with the exception of events where the source star falls within 0.5 kpc from the Galactic Nuclear Bulge where the uncertainty can get as large as 40%.

What is calculated here is the velocity of each point in the Milky Way Galaxy relative to the Galactic center. However, we need to find the relative transverse velocities with respect to a fixed observer, taken to be the planet Earth. This requires the introduction of two more parameters: the distance between the source and the Galactic center, and the angular separation ($\theta$) between the observer and the point source as viewed from the Galactic center.

Thus, assuming the observer to be standing on Earth, Equation (7) can be rewritten as:

$$v(r,\theta) = (25.62\ln(r) + 172.3) \times Cos\left(\theta + \arctan\left(\frac{r\sin\theta}{x_0 - r\cos\theta}\right)\right) \quad (8)$$

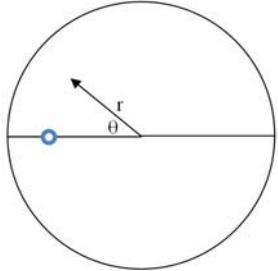

Figure 2 – As seen from the Galactic center, θ is the angular separation between the observer and the point source

Applying these velocities to the galactic disk for the Milky Way yields to the following figure 3:

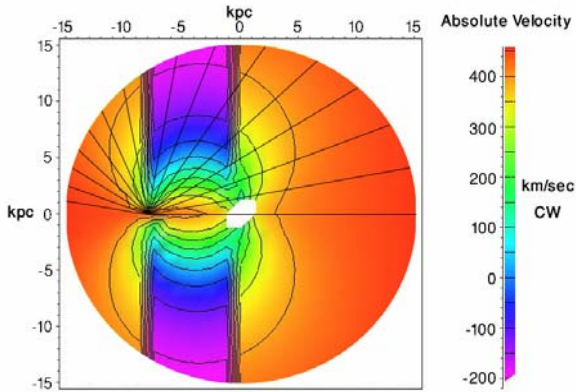

Figure 3 – Velocities of point sources relative to a stationary observer

Regions shown in red in the above sketch of our Galaxy show the highest velocity rates relative to Earth. On the other hand, regions shown in purple have negative velocities with respect to the line of sight, i.e. the line connecting the observer to any given source.

So far, the observer was assumed to be stationary and only absolute velocities were considered. However, since the Earth is also orbiting the center of the Galaxy, we may revise the above map to include the motion of the observer as well:

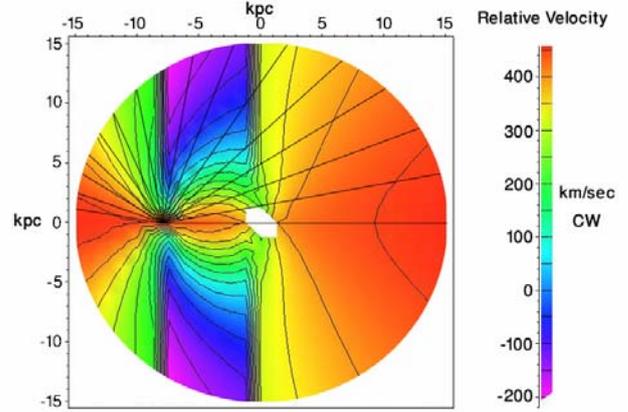

Figure 4 - Velocities of point sources relative to a moving observer

According to Equation (5), we need to independently provide a formula for each of the event lines, corresponding to the line of sight, in order to calculate the probability due to the relative velocities of the entire system, which makes calculations very complicated. To avoid such cumbersome computations, we may divide the Galactic Plane to 16 equal sections from the observer's point of view. The smaller these sections, the more accurate our result would be.

As illustrated in Figure (5), examining this latter graph (Figure 4) in three dimensions yields to an interesting illustration of our three dimensional Galaxy, showing a bridge-like structure connecting us to the world of extra-solar planets!

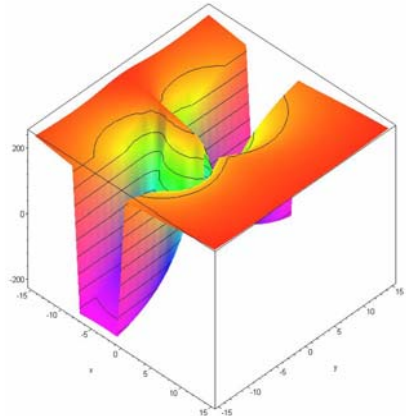

Figure 5 – A 3-D view of relative velocities shows a bridge-like structure in the middle

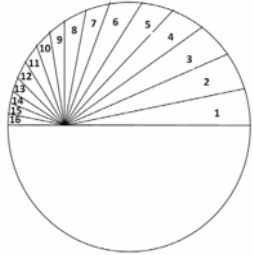

**Figure 6** – The upper half of the Galactic disk is divided into 16 sections, each having an 11.25 deg. Angular diameter as seen by an observer on earth

Here we again introduce two more parameters: sum of the weighted velocities, and sum of the weighted absolute values of velocities.

- Sum of the weighted velocities: $\sum Av$
- Sum of the weighted absolute values of velocities: $\sum A|v|$

where A is the area for each of the 16 sections introduced above.

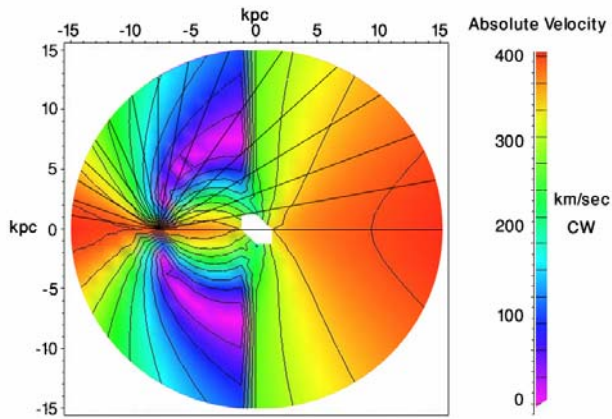

**Figure 7** – Absolute velocity

It is now easy to see how the event rate increases as the sum of the weighted velocities approaches to zero. This is due to each lens overlapping with each source as viewed from the Earth. On the other hand, this may be caused due to the smaller size of the section (such as section 16). To overcome this problem, we must determine the average velocities of each section independently, which is done easily using the sum of the weighted absolute values of the velocities. If the result turns out to be close to zero, this means the section under study is either small or the relative velocities of points within that section is almost zero.

The ratio of these two parameters $\left(\sum A|v|/\sum Av\right)$ would then help determine the relative velocity component of the event rate formula.

The graph below shows the event rate probability distribution based on relative velocities as viewed by the observer in his local coordinates. As this graph indicates, the probability in section 5 (corresponding to $45 < \theta < 55$) is the highest among others.

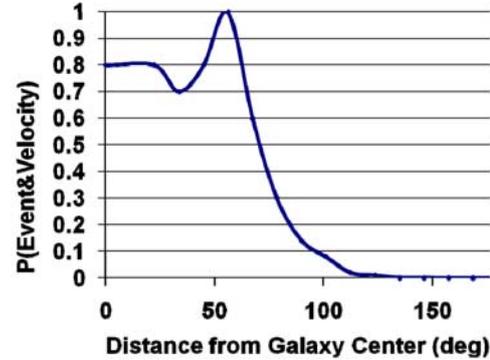

**Figure 8** – Probability distribution for the event rate based on relative velocities only

However, according to Equation (5), calculation of relative velocities is not adequate in finding the event rate in Microlensing. Two other parameters, namely the number of stars in the Microlensing Tube and the distribution of density along the line of sight, also affect the rate at which lenses enter the Tube.

### 2.2.2 DENSITY DISTRIBUTION

It is still not possible to accurately model the distribution of stars throughout the Galaxy. However, a number of different models have been presented in the past that are based on the most updated observational data. One of the simplest models provided is the exponential model that is attained based on the initial stellar density in the Galactic Nucleolus. The stellar distribution suggested by this model is independently applicable to any stellar type.

This enables us to derive a normalized graph showing the distribution of stars throughout the Galaxy. Here we have used dimensionless quantities and assumed the overall stellar distribution in the Galactic halo to be one.

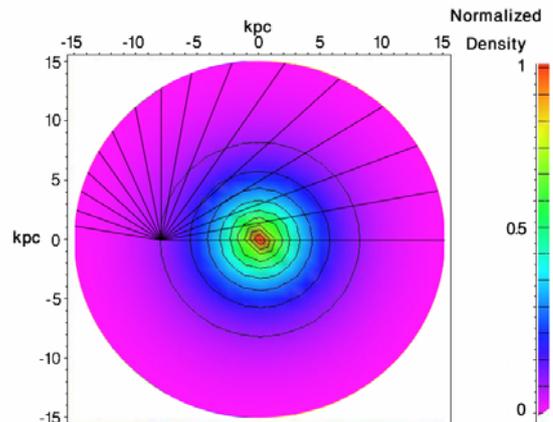

**Figure 9** – Normalized density distribution for the Milky Way Galaxy

As mentioned earlier, several other methods exist for modeling the stellar distribution. A combined use of these methods yields a more satisfactory result. Thus the probability distribution of Microlensing events can be inferred from stellar distribution that is generalized using linear density and integrating over the sections introduced earlier. This probability distribution is illustrated in Figure (10):

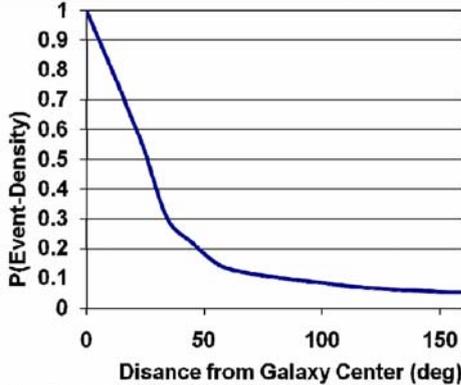

Figure 10 - Probability distribution for the event rate based on normalized densities only.

As expected, the event rate due to stellar distribution is higher at the Galactic center. On the other hand, we know that the two factors mentioned above, i.e. velocity and density, are independent properties and we must multiply them to get the overall probability. We may then scale down the graph to one, which is shown below:

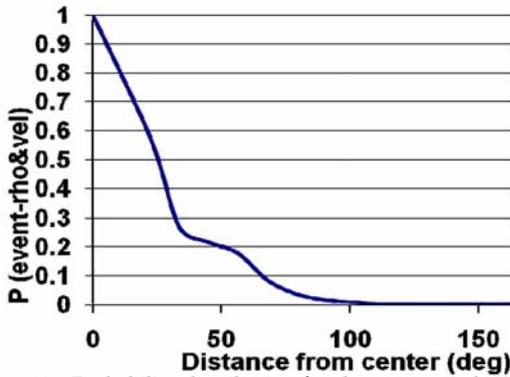

Figure 11 - Probability distribution for the event rate based on both relative velocities and normalized densities.

It is evident that the rate of decrease in the graph of Figure (11) momentarily seizes in the interval between 40 to 60 degrees from the center of the Galaxy. It seems the Galaxy is providing us the opportunity to detect planets in that region!!

## 3 SEARCH REGION AROUND TARGET STARS

For a star to host a habitable planet, it should be located within a region of the Galaxy that is known as the Galactic Habitable Zone. It should also be rich in metals in order to have rocky planet(s) around it that could potentially harbor life.

The graph in Figure (12) shows the universe's terrestrial planet production rate - the number of earths produced per year per cubic megaparsec. Earths have been produced since about 2.4 billion years after the Big Bang and our Earth was built 4.6 billion years ago, 8.8 billion years after the Universe was born.

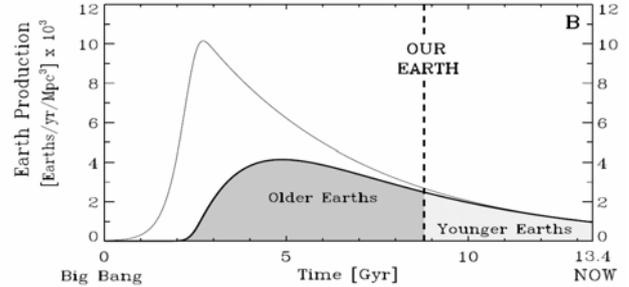

Figure 12 - Universe's terrestrial planet production rate (Courtesy of Lineweaver, 2001)

The dark grey area to the left of 8.8 billion years in Figure (12) is a measure of the number of earth-like planets older than ours, about $74\pm9\%$ are older. We live on a young planet. The first earth-like planets were formed about 11 billion years ago so the oldest are about 6.4 billion years older than our Earth. The age of the average earth in the Universe is $6.4\pm0.9$ billion years; that is, it formed about 7 billion years after the Big Bang. Thus, the average earth in the Universe is $1.8\pm0.9$ billion years older than our Earth. And, if life exists on some of these earths, it will have evolved, on average, 1.8 billion years longer than we have on Earth. For comparison, the thin line in the figure below is the star formation rate normalized to the earth production rate today. The time delay between the onset of star formation and the onset of earth production is the about 1.5 billion years that it took for metals to accumulate sufficiently to form earths (Lineweaver, 2001).

Therefore, in the search for such habitable earth-like planets, one must look for stars with high metallicity in the Galactic Habitable Zone.

### 3.1 GALACTIC HABITABLE ZONE

The stars most likely to host planets that could support life must be located within the Galactic Habitable Zone. This region is located 7-9 kpc from the galactic bulge and encompasses the entire Solar System. Two factors determine the boundaries of this zone. The inner (closest to the center of the Galaxy) limit is set by threats to complex life: nearby transient sources of ionizing radiation, including supernovae and gamma ray bursters, and comet impacts. Such threats tend to increase close to the Galactic center. The outer limit is imposed by galactic chemical evolution, specifically the abundance of heavier elements, such as carbon.

The probability of life evolving in our Galaxy as one moves away from the Galactic center is illustrated in Graph (13) below:

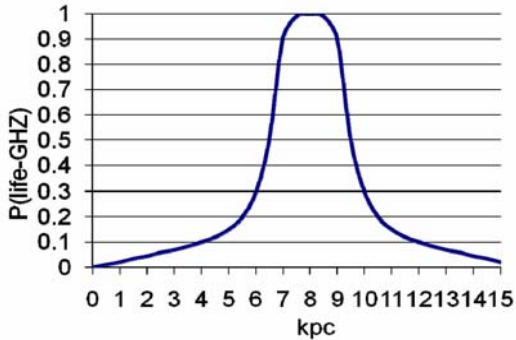

Figure 13 – The probability for life formation as a function of distance from the Galactic center.

The graph peaks in a range that falls between 7 to 9 kpc from the Galactic center with our Solar System sitting somewhere in the middle. Assuming the Galaxy to be a thin disk, we may obtain the schematic map below which is divided into equal sections as viewed by the observer, here taken to be the Earth:

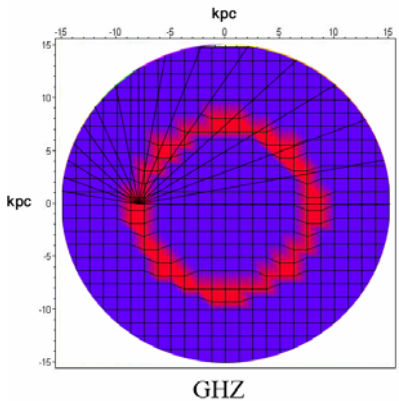

Figure 14 – The region shown in red marks the Galactic Habitable Zone

It is easy to see that the probability to detect habitable planets is equal in sections 1 through 4, peaks in section 5, and decreases as one moves to the last section where the probability reaches its lowest value. This probability distribution is illustrated in Figure (15):

The graph peaks at 60 deg. from the Galactic center. Surprisingly, this is consistent with the result obtained in Graph (11) earlier.

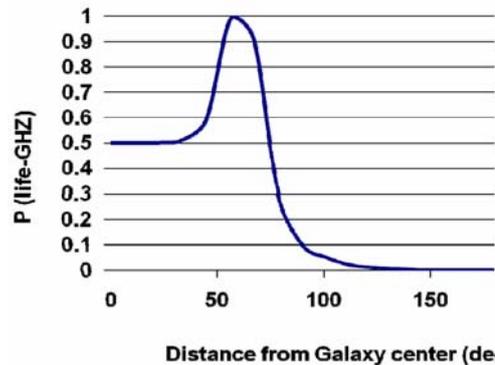

Figure 15 - Probability to detect habitable planets in the Milky Way Galaxy

## 3.2 METALLICITY

If metallicity had no effect on planet formation, we would expect the metallicity distribution of stars hosting Hot Jupiters (giant, close orbiting, extrasolar planets) shown in dark grey in Figure (12) to be an unbiased sub-sample of the distribution of sun-like stars in the solar neighborhood (light grey). However, Hot Jupiter hosts are more metal-rich. Hot Jupiters have the virtue of being Doppler-detectable but because they are so massive and so close to the host star and have probably migrated through the habitable zone, they destroy or preclude the existence of earths in the same stellar system.

Thus, the probability of destroying earths is the ratio of the dark histogram in Figure (16) to the light histogram. It is an estimate of the probability that a sun-like star of a given metallicity will have a Hot Jupiter. It is the ratio, as a function of metallicity, of the number of Hot Jupiter hosts to the number of stars surveyed.

The probability of harboring earths can be constrained by at least three considerations (Lineweaver, 2001):
1. At high metallicity, earths are destroyed or prevented from forming by the presence of Hot Jupiters,
2. At zero or very low metallicity, there are not enough metals to form earths,
3. Since the Earth and two other earth-like planets (Mars and Venus) exist around the Sun, it is plausible to suppose that terrestrial planets in general have a reasonable chance of forming around stars of near solar metallicity.

The probability of harboring earths shown in the figure below is based on the above-mentioned considerations and a production of earths that is linearly proportional to metallicity.

In this graph, the upper x-axis shows the Linear metal abundance. The Sun ($M \equiv Fe/H \equiv 0$) is more metal-rich than about 2/3 of local sun-like stars and less metal-rich than about 2/3 of the stars hosting Hot Jupiters. The high value of M (compared to neighboring stars) and the low value compared to Hot Jupiter hosts is expected if a strong metallicity selection effect exists.

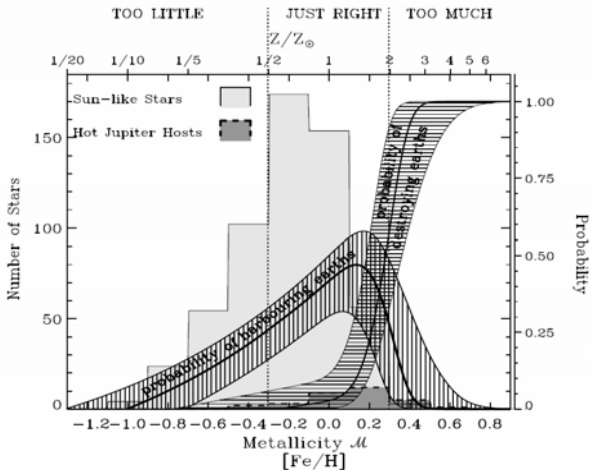

Figure 16 - The Metallicity Selection Effect (Courtesy of Lineweaver, 2001)

Thus determining "disaster zones" in a planetary system, regions that provide orbital instability for any planet located within, due to the presence of giant planets is also important in the search for habitable worlds. An earth-like planet located in this zone would either eventually come into collision with the larger planet, will hit the star, or will be knocked out of the system or at least move very far from its star. This requires having a full knowledge of the giant planets in a planetary system especially their mass and semi-major axis.

## 3.3 OTHER FACTORS TO CONSIDER

In addition to the aforementioned criteria, one needs to consider several other factors which define habitability for a planet. For instance, not only the entire stellar system must lie within the Galactic Habitable Zone, the planet itself must reside in the habitable zone of its host star which varies from one stellar system to another according to the mass and spectral type of the star harboring the planet. On the other hand, for intelligent life as we know it to form on a planet, the planet's eccentricity of orbit and axial tilt must lie within a certain range so that the change in the overall radiation received from the star would not overwhelmingly affect life on the planet.

There are of course several other criteria to examine when speaking of formation and evolution of life, especially intelligent life, on a planet, which we may skip for the time being to avoid diverting away from the objective of this paper.

## 4 RESULTS

### 4.1 Probability of Observing a Microlensing Event and Suggested Regions

We found earlier that habitability is closely linked to density distribution and velocity. It now becomes possible to find the overall probability distribution by multiplying these three independent factors, namely habitability, density, and velocity, which gives us the graph in Figure (17):

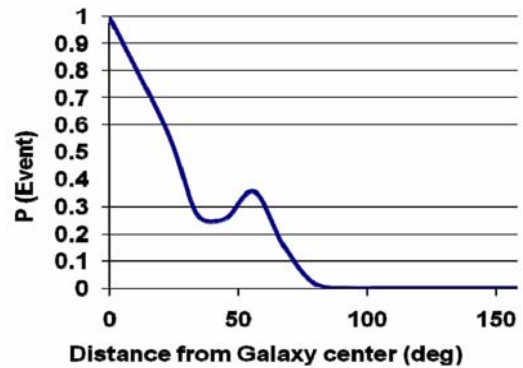

Figure 17 – Overall probability of detecting habitable planets

As mentioned earlier, galactic symmetry has enabled us to focus on half of the Galaxy and apply the results to the other half. Figure (18) puts this distribution on the celestial map, giving us regions with highest probability of detecting stars with habitable planets in Microlensing events.

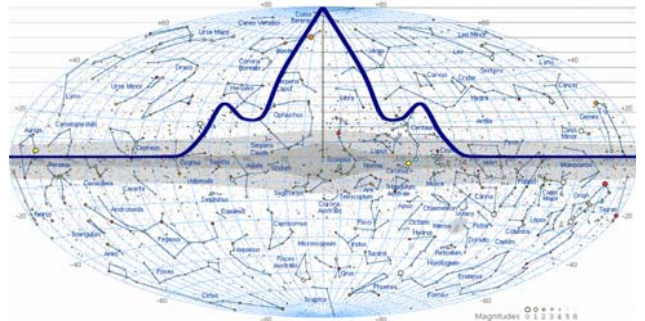

Figure 18 – Celestial map and probability on it.

The above map suggests three regions to be studied for Microlensing detection of extrasolar planets:
- o Galactic center
- o Toward Constellation Centaurus
- o Toward Constellation Sagittarius

Currently, observation of Microlensing events is done mostly toward either the Galactic center or the Magellanic Clouds through the Baade's Window. These two regions that are the targets of Microlensing studies are indeed located within the regions suggested by this work.

In order to observe a Lensing event by an intervening object or Microlensing by its accompanying planet throughout a month, one may easily estimate the number of stars needed to be monitored using mathematical calculations. This yields to about 100 million stars per month!

If we now calculate the area under the probability curve in Figure (17) and divide that number by 180 degrees, we obtain the mean probability distribution as viewed from any location on Earth. The number we get is approximately 0.1656 which again tells us that we should constantly monitor about 100 million starts to detect a habitable exoplanet. Thus we may draw a graph, shown as Figure (19), showing the number of stars needed to be observed over the course of one month to detect an extrasolar planet:

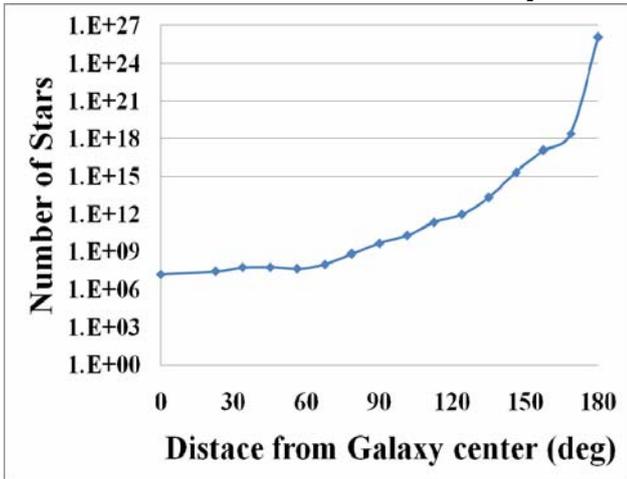

Figure 19 – This graph indicates that we need to monitor larger number of stars as we move farther

This shows that the chances to detect habitable exoplanets outside the region 60 degrees from the Galactic center is very slim with the current technology. Detection of such planets using ground-based observations is also limited by lack of sufficient facilities, as shown in Figure (20) below:

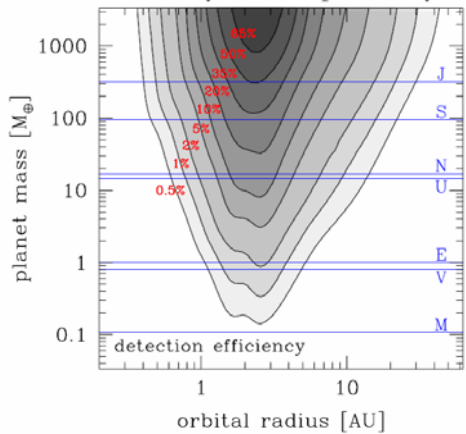

Figure 20 - The planet detection efficiency as a function of the orbital radius (assuming a circular orbit) and mass of the planet (Courtesy of M. Dominik et al., 2008)

Seeing conditions such as atmospheric turbulence, light pollution, and the observer's latitude as well as technological innovations such as the strength of optical detectors, and data processing methods are parameters which need to be taken into consideration when using ground-based equipments to detect exoplanets.

We may also examine the detection probability of an earth-like exoplanet using the ratio of distances between the lens and the source to the observer. We know that the strength of lensing increases as this ratio approaches 0.5, , assuming a fixed distance between the source and the observer, thus increasing the detection probability. Making note of the fact that the Einstein Radius is proportional to the square root of $x(1-x)$, as seen in Eq. (4), we may then draw another graph, shown in Figure (21) for a lens that is in motion along the observer's line of sight:

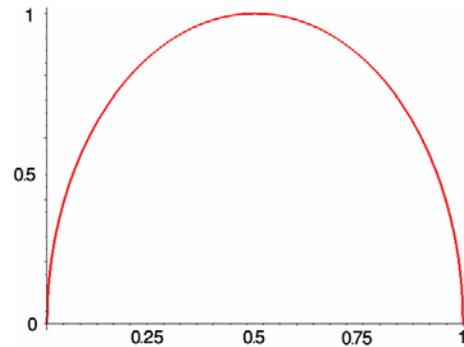

Figure 21 – Probability as a function of distance between the observer to the lens and source. The graph peaks precisely at the center from the Galactic center

Making note of this fact and the position of the Galactic Habitable Zone as shown in Figure (14), we may draw Figure (22) which suggests ideal positions for the sources: not concentrate on the first region in this work as a major part of the current research is focused on this area. However, the two other regions which fall on the two constellations Centaurus (to the right) and Sagittarius (to the left) are more of interest. These two regions are shown in Figures (23) and (24).

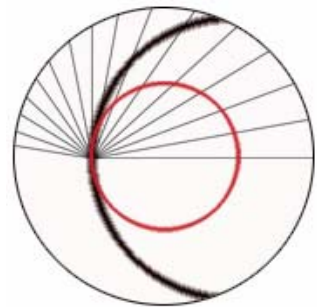

Figure 22 – The red circle is the Galactic Habitable Zone while the black curve shows positions ideal to be taken as sources to maximize probability of habitable planet detection. The radii of the two are in 1:2 ratio

Two major ideas can be inferred from this figure:

A.  We may rule out sections 9 through 16 since the detection probability decreases as the observer looks in the direction opposite the galactic bulge.

B. Considering stellar distribution throughout the Galaxy which yields more concentration of stars toward the galactic bulge as illustrated in Figure (9), it becomes evident that sections 1 through 4 have considerably less number of sources compared to sections 5 through 9.

Guided by the above ideas, two main patterns are suggested in the search for habitable exoplanets using the Planetary Microlensing Technique:
1. Detection when sources are abundant: This requires continuous observation to see a star passing in front of the source, which is the method used today.
2. Detection when lenses are abundant: This means concentrating on fewer stars.

## 4.2 SEARCH REGIONS

We found earlier from Figure (18) showing the probability distribution of Microlensing events that there are 3 regions in the galactic disk (assuming the mean thickness of the Galaxy to be 400 pc) with highest probability of finding habitable exoplanets: the Galactic center, seen through Baade's Window, and two more regions 45 to 55 degrees from the Glactic center. We may

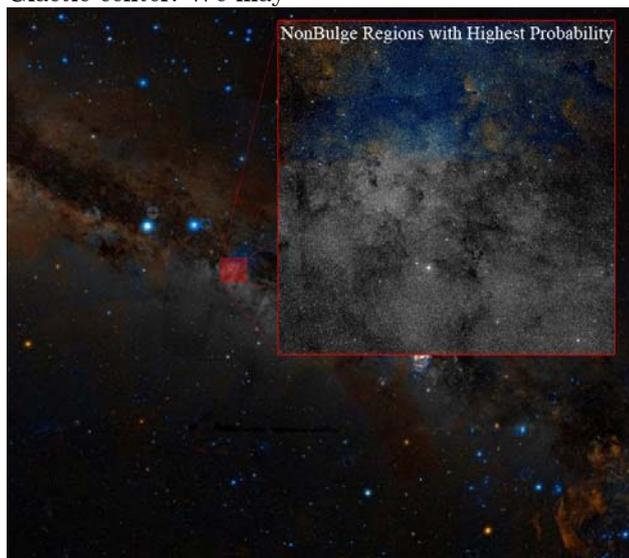

Figure 23 – Part of the Constellation Centaurus is one of the two regions suggested to search for habitable extrasolar planets

The region under study in Centaurus turns out to be illuminated from background sources which make detection easier. However, Sagittarius seems to be dark which is due to concentration of interstellar gas and dust, making it less appropriate for detecting exoplanets. However, this is only when we scan these two regions using the visible part of the electromagnetic spectrum which indeed covers a very small portion.

Thus we need to examine these two regions using other parts of the spectrum to come up with a more reasonable result.

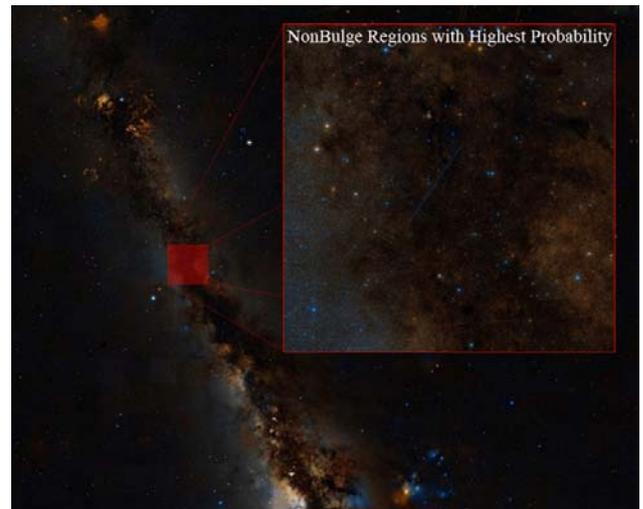

Figure 24 – Part of Constellation Sagittarius is one of the two regions suggested to search for habitable extrasolar planets

## 4.3 Observation in Different Parts of the Electromagnetic Spectrum

Since observation of Sagittarius in visible wavelengths is blurred, we may use different wavebands to find out which one is more efficient in our study. This is shown in Figure (25).

In order to detect exoplanets through the Microlensing technique in the two regions suggested above, it now becomes evident that using wavelengths of atomic hydrogen and molecular hydrogen would lead us to finding more of such planets.

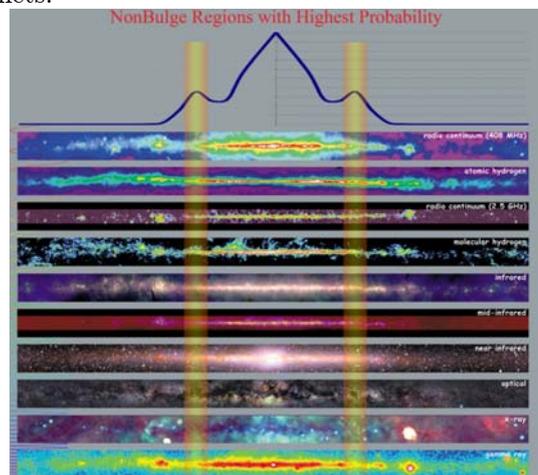

Figure 25 – Studying regions with highest probability of habitable exoplanet detection in Centaurus and Sagittarius through a range of different wavelengths (NASA).

# 5 SCIENTIFIC RECOMMENDATIONS

In the search for habitable worlds like ours beyond the limits of the Solar System using the Gravitational Microlensing Technique, the following scientific recommendations are proposed:

1. In addition to the Galactic center where most of the current search is focused on, we may look in two regions located in constellations Centaurus and Sagittarius where the probability to detect habitable worlds is high;

2. To come up with a more accurate map showing regions with highest probability of exoplanet detection, it is necessary to enhance our understanding of distribution of stars of various spectral types and their velocities throughout the Galaxy;

3. It turns out that observing regions where lenses are in abundance seems to be more appropriate as it increases the chances of detecting more planets. Thus, it is recommended to concentrate our studies on regions where one might find large numbers of lenses instead of focusing on abundant sources;

4. The 21.1-cm line atomic hydrogen is preferred over other portions of the spectrum in detecting habitable exoplanets;

5. Although planetary Microlensing can find small planets, it is a one-time event and is essentially non-repeatable, which suggests a collaboration of activities to catch a planet on the run.

6. Throughout the formulation and development of this map, it is important to develop the scientific community to make use of as well as refine and prioritize the target list to identify stars most likely to harbor earth-like planets.

# 6 CONCLUSION

One of the most compelling questions that modern science can address is whether or not earth-like planets, habitable or already life-bearing, exist elsewhere in the Universe. Thus in addition to understanding the formation and evolution of planets and, ultimately, of life beyond our Solar System, one must learn where such planets are more likely to be found.

This document presented a distribution map for probability of finding stars within the Milky Way Galaxy that could harbor terrestrial planets capable of hosting life, with a particular emphasis on the Microlensing detection of exoplanets. With the planet count now standing at over 405 while only nine have been discovered using the Gravitational Microlensing Technique as of 13 November 2009, we predict that the result of this work would boost this number in the not-so distant future as it could speed up the pre-design phase for space missions and provide more promising target stars for ground and space based observations.

Thus, in this document, considerable emphasis was placed on performing a thorough, systematic, and comprehensive study of potential Microlensing targets that can be detected by observations using ground-based and space-based observatories. Nonetheless, as our understanding of our Galaxy improves, the current map will be refined to include other regions where detection of a planet through Microlensing is also possible.

Since Microlensing events are rare, the observing strategy is to monitor large numbers of stars in regions of sky in the direction of constellations Centaurus and Sagittarius, using the 21.1-cm atomic hydrogen, in addition to the already practiced strategy which places the focus on the Galactic Bulge.

Also we are going to do a strong review on velocity curve of Galaxy stars, Density of Stars and Dust, Instrument limitations and adding caustic features in our future model for better accuracy.